\documentclass[prb,twocolumn,superscriptaddress]{revtex4-2}
\usepackage{ulem}
\usepackage{graphicx,amsmath,amssymb,amsfonts,verbatim,dsfont,color}
\usepackage{bbm, bm, latexsym}
\usepackage{hyperref}
\usepackage[us]{datetime}
\usepackage{tikz,soul}
\usepackage{lmodern}

\graphicspath{ {pics/} }

\definecolor{lime}{HTML}{A6CE39}
\DeclareRobustCommand{\orcidicon}{%
	\begin{tikzpicture}
	\draw[lime, fill=lime] (0,0)
	circle [radius=0.16]
	node[white] {{\fontfamily{qag}\selectfont \tiny ID}};
	node[white] {{\fontfamily{cmr}\selectfont \tiny ID}}
	\draw[white, fill=white] (-0.0625,0.095)
	circle [radius=0.007];
	\end{tikzpicture}
	\hspace{-2mm}
}

\foreach \x in {A, ..., Z}{%
	\expandafter\xdef\csname orcid\x\endcsname{\noexpand\href{https://orcid.org/\csname orcidauthor\x\endcsname}{\noexpand\orcidicon}}
}

\newcommand{\kp}{$k\cdot p$ }

\newcommand{\thetitle}{Optical control of conductivity type and valley polarization via persistent photoconductivity in (Pb,Sn)Se quantum wells}

\begin{document}

\title{\thetitle}

\author{Alexander~Kazakov\orcidA}
\email{kazakov@MagTop.ifpan.edu.pl}
\affiliation{International Research Centre MagTop, Institute of Physics, Polish Academy of Sciences, Aleja Lotnikow 32/46, PL-02668 Warsaw, Poland}
\author{Gauthier~Krizman\orcidB}
\affiliation{Laboratoire de Physique de l’Ecole normale sup{\'e}rieure, ENS, Universit{\'e} PSL, CNRS, Sorbonne Universit{\'e}, 24 rue Lhomond 75005 Paris, France}
\author{Valentine~V.~Volobuev\orcidH}
\affiliation{International Research Centre MagTop, Institute of Physics, Polish Academy of Sciences, Aleja Lotnikow 32/46, PL-02668 Warsaw, Poland}
\affiliation{National Technical University "KhPI", Kyrpychova Str. 2, 61002 Kharkiv, Ukraine}
\author{Micha\l~Szot\orcidJ}
\affiliation{International Research Centre MagTop, Institute of Physics, Polish Academy of Sciences, Aleja Lotnikow 32/46, PL-02668 Warsaw, Poland}
\author{Wojciech~Wo{\l}kanowicz\orcidK}
\affiliation{International Research Centre MagTop, Institute of Physics, Polish Academy of Sciences, Aleja Lotnikow 32/46, PL-02668 Warsaw, Poland}
\author{Chang-Woo~Cho\orcidC}
\affiliation{Laboratoire National des Champs Magn{\'e}tiques Intenses, CNRS, LNCMI, Universit{\'e} Grenoble Alpes, Univ. Toulouse, INSA Toulouse, EMFL, F-38042 Grenoble, France}
\affiliation{Department of Physics, Chungnam National University, Daejeon, 34134, Republic of Korea}
\affiliation{Institute for Sciences of the Universe, Chungnam National University, Daejeon 34134, South Korea}
\author{Benjamin~A.~Piot\orcidD}
\affiliation{Laboratoire National des Champs Magn{\'e}tiques Intenses, CNRS, LNCMI, Universit{\'e} Grenoble Alpes, Univ. Toulouse, INSA Toulouse, EMFL, F-38042 Grenoble, France}
\author{Tomasz~Wojciechowski\orcidE}
\affiliation{International Research Centre MagTop, Institute of Physics, Polish Academy of Sciences, Aleja Lotnikow 32/46, PL-02668 Warsaw, Poland}
\author{Gunther~Springholz\orcidF}
\affiliation{Institut f{\"u}r Halbleiter- und Festk{\"o}rperphysik, Johannes Kepler University, Altenbergerstrasse 69, A-4040 Linz, Austria}
\author{Tomasz~Wojtowicz\orcidG}
\affiliation{International Research Centre MagTop, Institute of Physics, Polish Academy of Sciences, Aleja Lotnikow 32/46, PL-02668 Warsaw, Poland}
\author{Tomasz~Dietl\orcidI}
\email{dietl@MagTop.ifpan.edu.pl}
\affiliation{International Research Centre MagTop, Institute of Physics, Polish Academy of Sciences, Aleja Lotnikow 32/46, PL-02668 Warsaw, Poland}

\begin{abstract}
The ability to tune the Fermi level of semiconductors is at the heart of modern electronics. Here, we demonstrate that persistent photoconductivity (PPC) enables tuning of carrier density, conductivity type, and, consequently, the valley polarization in (Pb,Sn)Se/(Pb,Eu)Se quantum wells. Illumination of these samples induces Fermi level shifts that convert the system from a threefold-degenerate $\bar{M}$-valley two-dimensional hole gas to a single $\bar{\Gamma}$-valley-polarized electron gas with similar values of mobility. The optically induced state persists for more than $10^{3}$ minutes at cryogenic temperatures and enables stepwise optical gating without the need for device processing. These transitions are confirmed by the sign inversion of the Hall slope and the modification of quantum Hall plateau degeneracies measured in magnetic fields up to 35~T. Landau level \kp model calculations quantitatively reproduce the experimental data. Furthermore, studies of weak-field magnetoresistance demonstrate the significance of quantum localization phenomena at the transition between the weakly and strongly localized regimes in compensated narrow-gap semiconductors. Spectral studies allow us to identify the critical role of the barrier material and determine the photon energies that can reverse the PPC effect. The persistent light-induced upward shift of the Fermi level in the $p$-type quantum well is explained in terms of specific energy positions of donor and acceptor defect states in the studied system. Our results demonstrate that PPC is a powerful optical gating tool for the IV-VI quantum wells, a versatile platform for reconfigurable valleytronic architectures. Furthermore, we discuss possible mechanisms allowing for fine, persistent, and reversible tuning by illumination between hole and electron conductivity and the relevance of our results for realizing the quantum spin Hall and quantum anomalous Hall effects in topological crystalline insulators.
\end{abstract}

\date{\today}
\maketitle

\section{Introduction}

Controlling the charge carrier density — and hence the position of the Fermi level $E_{\mathrm{F}}$ --- remains a central challenge in the study of topological edge transport and, more broadly, in the investigations of electronic band structures. To detect the quantum spin Hall effect (QSHE) \cite{Bernevig2006,Koenig2007}, or quantum anomalous Hall effect (QAHE) \cite{Yu2010,Chang2013}, for instance, $E_{\mathrm{F}}$ must reside within the bulk bandgap. Only then can robust, quantized edge-state conductivity at zero magnetic field be demonstrated; otherwise, hybridization between edge and bulk states obscures the topological signatures. Carrier density $n$ is conventionally tuned through electrostatic gating. Yet, an alternative and conceptually distinct approach exploits the persistent photoconductivity (PPC) effect \cite{Sheinkman1976,Sumanth2022}: a photoinduced modification of conductivity that endures long after the illumination is turned off. This phenomenon has been observed in a wide variety of systems, ranging from bulk semiconductors \cite{Semaltianos1995,Akimov2000} to heterostructures \cite{Knebl2018,Bolanos2022,Sotnichuk2025} and atomically thin materials \cite{DiBartolomeo2017,Grillo2021}. Of particular interest is bipolar PPC \cite{Yeats2015,Nikolaev2022}, in which illumination of different photon energies $\hbar\omega$ can either increase or decrease $n$, offering a purely optical route to $E_{\mathrm{F}}$ control.

The PPC phenomenon has been extensively studied in both wide- and narrow-gap semiconductors, serving as a powerful tool in fundamental transport research \cite{Samani2014,Kazakov2017} and as a functional mechanism in optoelectronic devices \cite{Sumanth2022}. In particular, by controlling the Fermi level position, PPC also finds its application while studying topological systems \cite{Yeats2015,Nikolaev2022}, potentially leading to conditions allowing the observation of topological edge transport signatures, such as QSHE or QAHE. While these effects have been realized in several material platforms \cite{Koenig2007,Wu2018}, they have not yet been observed in an important class of topological matter—--topological crystalline insulators (TCIs). Distinguished from time-reversal-protected topological insulators, TCIs derive their nontrivial topology from crystal point-group symmetries and are realized in rock-salt IV–VI narrow-gap semiconductors under appropriate conditions \cite{Fu2011,Hsieh2012,Dziawa2012,Tanaka2012,Xu2012}. In these systems, the crystal orientation plays a decisive role in determining the nature of the surface states \cite{Liu2013a,Safaei2015,Liu2015}. Epitaxial IV–VI quantum well (QW) heterostructures are most commonly grown along the $\langle111\rangle$ direction, employing Eu-alloyed barriers \cite{Springholz2013}. Such architecture provides almost symmetric type I band alignment in the (Pb,Sn)Se/(Pb,Eu)Se QWs \cite{Simma2012,Krizman2018}. Their band structure gives rise to two distinct types of valleys: a single longitudinal valley at $\bar{\Gamma}$ and a triply degenerate oblique valley at $\bar{M}$, which are energetically split by biaxial strain imposed by the barrier material \cite{Yuan1994,Krizman2024}. Realization of the topological phases in IV-VI semiconductors requires an increased Sn content, which inevitably leads to the formation of a cation vacancy that acts as a double acceptor, thereby increasing the hole concentration \cite{Nimtz1983}. This fact makes control of the Fermi energy $E_{\mathrm{F}}$ crucial for exploring topological phenomena such as QSHE and QAHE, theoretically predicted to exist in specific quantum structures of IV-VI compounds \cite{Safaei2015,Liu2015,Kazakov2025,Majewski2026}. In recent years, there is increased interest in IV-VI compounds also in the context of other quantum technologies \cite{Kate2022,Gomanko2022,Schellingerhout2023,Song2023,Hussain2024a,Cuono2025}, topological properties \cite{Mazur2019,Brzezicki2019,Nguyen2022,Hussain2024,Kawala2025}, and valleytronics \cite{Krizman2024,Yoshimi2025,Krizman2025}, where the control of the $E_{\mathrm{F}}$ is also important.

\begin{figure}
    \centering
    \includegraphics[width=1.00\columnwidth]{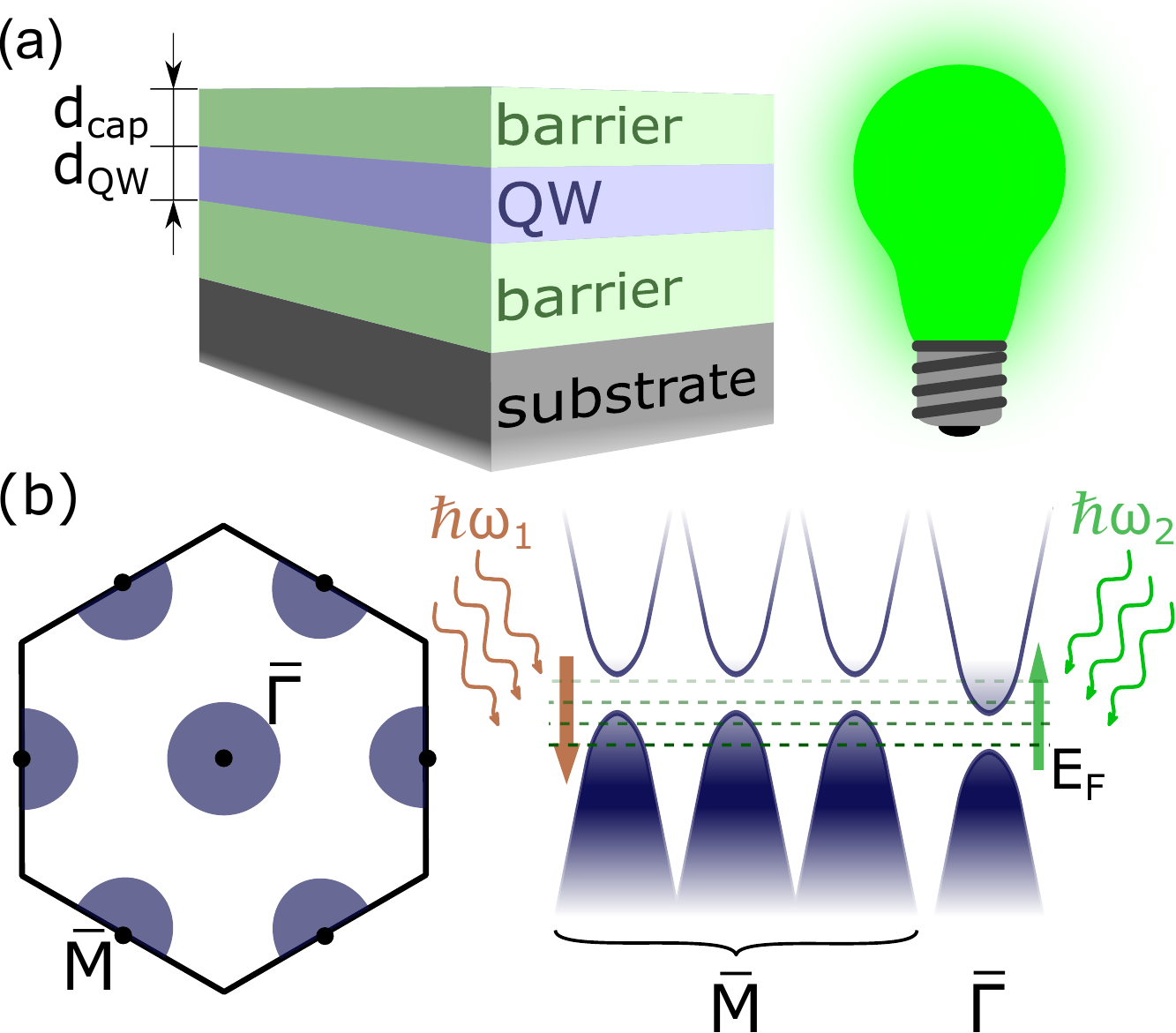}
    \caption{
    (a) Schematic of the experiment: green LED (on the right) is placed close to the (111)-oriented (Pb,Sn)Se/(Pb,Eu)Se QW heterostructure (on the left). 
    (b) Schematically depicted 2D projection of the Brillouin zone, with bands at $\bar{\Gamma}$ and three inequivalent $\bar{M}$ points. Illumination of the QW with photons effectively changes the Fermi level $E_{\mathrm{F}}$ of the system. Due to the biaxial strain, valleys in the $\bar{M}$ and $\bar{\Gamma}$ are shifted on the energy scale relative to each other. Thus, illumination causes the system to switch between 3-valley degenerate 2DHG and 1-valley polarized 2DEG. Different values of photon energies cause the shift of the Fermi level $E_{\mathrm{F}}$ either up (case of $\hbar\omega_{2}$) or down (case of $\hbar\omega_{1}$).
    }
    \label{fig:bands}
\end{figure}

Studies of PPC in such heterostructures have focused predominantly on Te-based QWs \cite{Pena2017,Lopes2021,Morais2021,Bolanos2022}, with recent evidence extending this behavior to Se-based systems \cite{Kazakov2025}. Up to now, there was no evidence for the optically induced persistent {\it p-n} transition in IV-VI quantum structures. In this work, we uncover a pronounced and remarkably long-lived PPC in (Pb,Sn)Se/(Pb,Eu)Se QWs, manifested as a strong photoresponse at low temperatures that persists for several days after the illumination is discontinued. Samples measured in darkness display transport characteristics consistent with previous investigations on the same structures \cite{Krizman2024,Kazakov2025}. We focus on two principal aspects of this phenomenon. First, we demonstrate its practical potential by showing that green light-emitting diode (LED) illumination: (i) enables efficient tuning of carrier density, quantum Hall states, and valley polarization (Fig.~\ref{fig:bands}a) and (ii) sheds a new light on the nature of the transition between weakly and strongly localized regimes. Second, we examine the spectral response of QW conductivity to elucidate the microscopic mechanisms underlying the PPC effect. While exploring it, we find that certain photon energies can reverse the PPC, allowing for full control over the $E_{\mathrm{F}}$ (Fig.~\ref{fig:bands}b). We explain the light-induced persistent upward shift of the Fermi level in the $p$-type QW to the asymmetry in the location of the acceptor and donor defect states with respect to band edges in (Pb,Sn)Se and (Pb,Eu)Se. Taken together, the results obtained in our study establish PPC as a versatile, contactless means of manipulating quantum transport in IV–VI topological structures.

\section{Samples and experimental details}

The studied (Pb,Sn)Se/(Pb,Eu)Se QWs were grown on (111)-oriented BaF$_2$ substrates by molecular beam epitaxy (MBE) employing the same Eu content in the buffer and cap layers, as described in previous works \cite{Krizman2024,Kazakov2025}. The series comprises four samples with different Sn contents, as summarized in Table~\ref{tab:info}. The first sample, denoted A, was Sn-free and exhibited insulating behavior at low temperatures before illumination. Samples B and C displayed {\it p}-type conductivity and possessed nearly vanishing band gaps, while sample D was heavily {\it p}-type, with an Sn concentration sufficient to invert the bulk band gap. Although the compositions cover a wide range of Sn content, none of the QWs met the conditions required to host the quantum spin Hall phase \cite{Kazakov2025}. The discussion in the main text focuses primarily on the unprocessed sample B. Other samples, wet-etched into a Hall-bar shape, exhibit qualitatively similar behavior (see Supplemental Material (SM) \cite{sup}), which provides evidence that the effect is sample-independent and that processing does not destroy it.

The majority of magnetotransport studies were performed in a $^3$He cryostat in magnetic fields up to 9~T. Measurements up to 35~T were performed using a dc resistive magnet and dilution refrigerator at the Grenoble high magnetic field facility. Illumination was provided by a green LED (with $\lambda$ = 550~nm and FWHM = 5~nm at T = 4.2~K) positioned to produce diffusely scattered light, ensuring good uniformity of the sample surface exposure. The standard lock-in technique was used to measure resistance, employing an ac excitation current of 100~nA.

Spectral characteristics of the PPC effect were measured in a Janis ST-100 optical cryostat equipped with electrical feedthroughs, which allows for sample temperature control within the range of 4–-300~K. The samples were cooled in complete darkness and subsequently illuminated with light of a specified wavelength. For measurements in the (400--800)~nm wavelength range, a xenon lamp and a double-grating monochromator of a Jobin Yvon FL3-2iHR spectrometer were used, enabling selection of light wavelength with a resolution of approximately 0.5 nm. In the wavelength range from 800~nm to 4~$\mu$m, a Newport 6363 IR infrared light source and a Jobin Yvon IHR 320 monochromator were employed. The intensity of the light incident on the sample was controlled using optical filters and the monochromator slit. Beam size ensured that the sample surface was uniformly exposed. The spectral characteristics of the light after passing through the optical measurement path were verified using a Thorlabs PM320E dual-channel optical power and energy meter, with its sensors placed at the exit of monochromators.

\begin{table}
\caption{Information about the QWs employed in a current study, i.e. nominal Sn and Eu content ($x_{\mathrm{Sn}}$ and $y_{\mathrm{Eu}}$), QW thickness $d_{\mathrm{QW}}$, cap layer thickness $d_{\mathrm{cap}}$, barrier material nominal band gap $E_{\mathrm{g}}^{\mathrm{barrier}}$, dark sheet carrier density $n_{\mathrm{2D}}$ ('-' sign indicate holes) and dark carrier mobility $\mu$.}
\begin{tabular}{c|c c c c c c c}
\hline
QW & $x_{\mathrm{Sn}}$ & $y_{\mathrm{Eu}}$ & $d_{\mathrm{QW}}$ & $d_{\mathrm{cap}}$ & $E_{\mathrm{g}}^{\mathrm{barrier}}$ & $n^{\mathrm{dark}}_{\mathrm{2D}}$, & $\mu^{\mathrm{dark}}$\\
 & \% & \% & nm & nm & meV & $10^{12}$~cm$^{-2}$ & cm$^2$/Vs\\
\hline
A & 0 & 6.5 & 25 & 200 & 341 & -- & --\\ 
B & 11 & 8.5 & 25 & 300 & 401 & -0.75 & 38000\\ 
C & 16 & 6.5 & 25 & 200 & 341 & -1.15 & 150000\\ 
D & 25 & 9.5 & 10 & 200 & 431 & -3.15 & 5000\\ 
\hline
\end{tabular}
\label{tab:info}
\end{table}

\section{Results and Discussion}

\subsection{Impact of illumination on conduction}

\begin{figure}
    \centering
    \includegraphics[width=1.00\columnwidth]{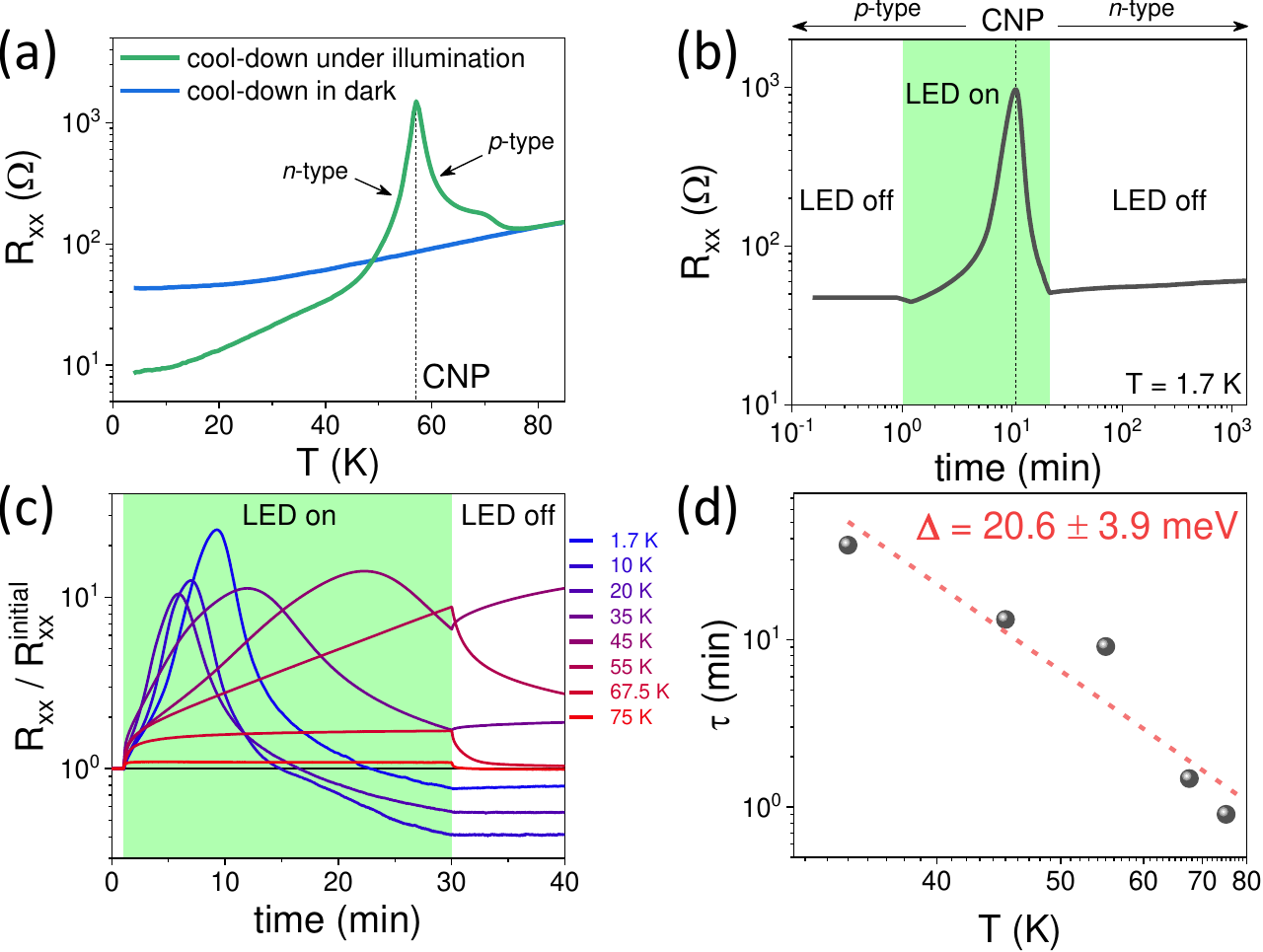}
    \caption{
    (a) Temperature dependence of the longitudinal resistance $R_{\mathrm{xx}}$ measured in darkness and under illumination. The divergence between the two traces grows upon cooling, with the resistance under illumination exhibiting a pronounced peak at low temperature.
    (b) Time evolution of $R_{\mathrm{xx}}$ at 1.7~K, demonstrating the persistent photoconductivity effect. When the LED is switched on ($I_{\mathrm{LED}} = 20~\mu$A), the resistance rapidly evolves through the peak exhibiting first negative, and then positive photoconductivity. After the light is turned off, it remains in a metastable state for more than a day.
    (c) Temperature dependence of the PPC dynamics. With increasing temperature, the photoresponse slows down, while the relaxation following illumination shutdown becomes progressively faster.
    (d) Exponential fit to the post-illumination decay yields the relaxation time $\tau$, which increases exponentially with decreasing temperature.}
    \label{fig:relax}
\end{figure}

The effect of illumination becomes already apparent from the comparison of the temperature dependence of resistance without and in the presence of illumination. In the absence of illumination, sample B exhibits metallic behavior at low temperatures: the longitudinal resistance $R_{\mathrm{xx}}$ decreases upon cooling, as shown in Fig.~\ref{fig:relax}a. However, the behavior drastically changes under illumination conditions. When the sample is cooled while exposed to light from the LED, the $R_{\mathrm{xx}}$ initially coincides with the 'dark' value but begins to diverge below $\approx$75~K. The 'light' trace then develops a pronounced peak, with a maximum over an order of magnitude higher than that of the 'dark' $R_{\mathrm{xx}}$ value at the same temperature.

This observation shows that the studied QWs have a strong electrical response to the applied light. When a sample cooled in the dark is subsequently illuminated at low temperature (Fig.~\ref{fig:relax}b), a similar resistance peak emerges as a function of time. After $\approx$20~min of illumination, when the LED is switched off, the resistance does not return to its initial value. This behavior constitutes clear evidence of persistent photoconductivity (PPC). As confirmed by magnetotransport measurements (discussed below), the effect originates from photoinduced electron doping of the QW, and the peak corresponds to the electron-hole compensation, in other words, to the charge neutrality point (CNP). The photoinduced state remains stable for more than 17~hours, demonstrating that illumination serves as an efficient and long-lived optical gating tool.

The effect is strongly temperature dependent. Measurements performed at different temperatures show that the photoresponse slows down as temperature increases, mainly reflecting the competition between photoinduced carrier generation and thermally activated recombination processes (Fig.~\ref{fig:relax}c). Partially, the temperature dependence is due to the change in the LED intensity with heating. Already above $\approx$65~K, recombination becomes sufficiently fast that the two-dimensional system returns to its initial state within a few minutes. Fitting the post-illumination decay of resistance with an exponential function yields characteristic relaxation times $\tau$ of several tens of minutes at 35~K (Fig.~\ref{fig:relax}d). At lower temperatures, relaxation becomes so slow that it cannot be reliably extracted within the chosen measurement protocol.

The temperature dependence of the relaxation time, $\tau(T)$, follows an activated behavior, $\tau \propto \exp(-\Delta/k_{\mathrm{B}}T)$, with an activation energy of $\Delta \approx 20$~meV ($\sim240$~K) (Fig.~\ref{fig:relax}d). This characteristic energy scale is of the order of the characteristic temperature at which the light-induced resistance change becomes visible. Depending on the sample, the divergence occurs at temperatures of approximately 40–-80~K \cite{sup}, corresponding to thermal energies of 3.4–-6.9~meV. Importantly, this appears to be primarily due to the non-zero conductivity of the barrier material at high temperatures. While measuring the conductivity of the barrier material, we observed thermally activated conductivity that persists down to temperatures of about 80~K \cite{sup}. As discussed in Sec~\ref{sec:mechanism}, above this temperature, holes trapped by acceptors can migrate toward the interface, thereby facilitating full recombination between the conducting electrons accumulated in the QW and the localized positive charges in the barriers.

\subsection{Tuning Quantum Transport via Persistent Photoconductivity}


We now turn to the influence of PPC on the low-temperature magnetotransport properties of the QWs. The long values of $\tau$ at low temperatures make it possible to controllably shift the Fermi level, $E_{\mathrm{F}}$. In this approach, the LED acts as a global optical non-contact gate that tunes the carrier density $n$. In the experiments described below, a current of 20~$\mu$A was passed through the LED for 30~s; after switching off the illumination, the two-dimensional system was allowed to relax for 10~min before magnetoresistance measurements. Repeated illumination cycles incrementally shifted $E_{\mathrm{F}}$, enabling magnetotransport measurements across a sequence of well-defined carrier doping states.

\begin{figure}
    \centering
    \includegraphics[width=1.00\columnwidth]{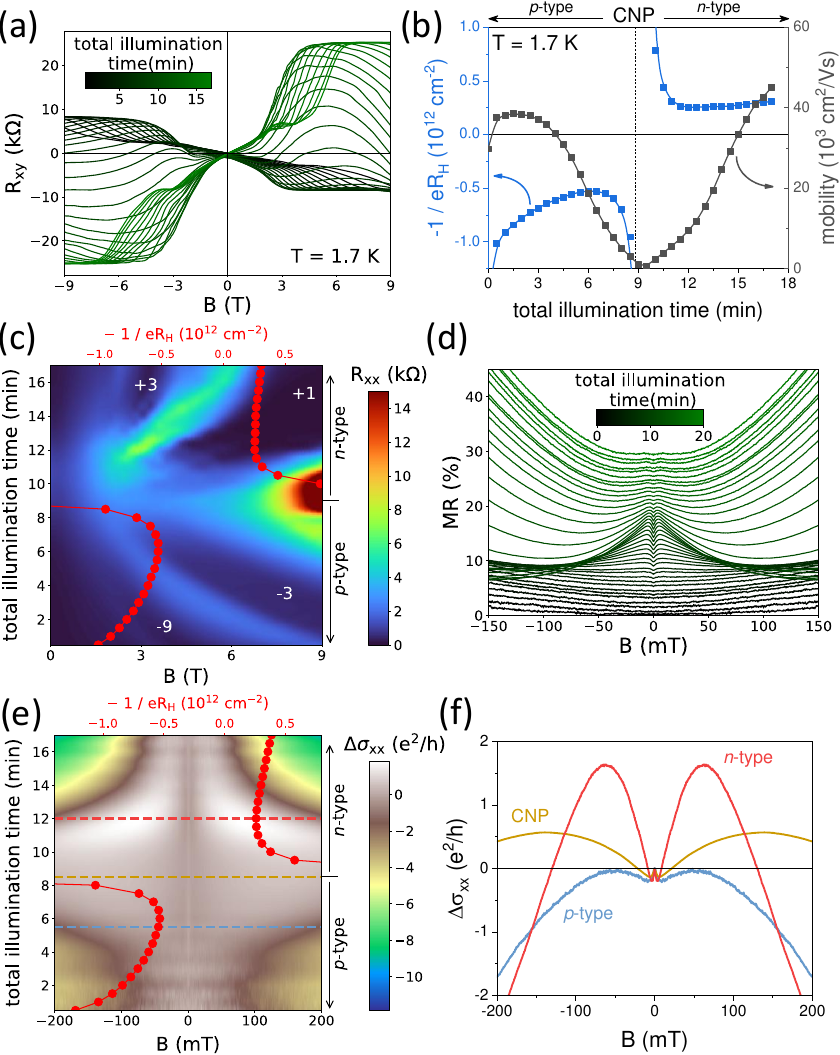}
    \caption{
    (a) The evolution of the Hall resistance as a function of cumulative illumination time. Each curve corresponds to a magnetotransport measurement performed after a short illumination period, revealing a systematic change of the Hall slope from hole-like to electron-like. 
    (b) Extracted carrier density ($n$) and Hall mobility ($\mu$) as functions of total illumination time. The sign change of $n$ (with zero marked by a black horizontal line) confirms a transition through the charge neutrality point.    
    (c) High-field magnetoresistance (MR) showing the evolution of Shubnikov–de Haas oscillations. The annotated filling factors $\nu$ are labelled in white, with ‘$+$’ and ‘$-$’ indicating electron-like and hole-like regimes, respectively.
    (d) Low-field part of magnetoresistance showing non-monotonic behavior consistent with weak antilocalization (WAL) and weak localization (WL). Each curve is offset by 1.5\% for clarity.
    (e) Low-field magnetoconductivity demonstrating WAL and WL near the charge neutrality point. In panels (c) and (e), red markers indicate carrier densities corresponding to the upper-axis values, determined from the total illumination time.
    (f) Representative WL-WAL behavior in {\it p}- and {\it n}-type regimes and in the CNP, colors of the curves correspond to those of the dashed lines in (d).
    All measurements in (a-f) were performed at 1.7~K.}
    \label{fig:RvsB}
\end{figure}

Initially, the Hall resistance $R_{\mathrm{xy}}$ exhibits a negative slope (Fig.~\ref{fig:RvsB}a), characteristic of hole-like transport. With further increasing illumination dose, the Hall resistance gradually evolves and eventually acquires a positive slope, signalling a transition to electron-like transport. Here we use the convention that $R_{\mathrm{xy}}$ has a positive slope for electrons, while the Hall coefficient $R_{\mathrm{H}}$ is negative (for holes, signs are opposite). We extract the inverse Hall coefficient $1/eR_{\mathrm{H}}$ from the low-field Hall slope and estimate the Hall mobility $\mu$ from the zero-field resistance (Fig.~\ref{fig:RvsB}b). At certain photon doses, we achieve compensation between electrons and holes, i.e. CNP, where $1/eR_{\mathrm{H}}$ diverges and $\mu$ approaches zero.

In both the {\it n}- and {\it p}-type regimes, $\mu$ exhibits a similar trend: it increases monotonically with the carrier density and ultimately reaches comparable values. Data for other samples \cite{sup}, follow the same trend, except for sample D, where $\mu$ increases as the hole density decreases. Earlier studies on bulk IV–VI semiconductors have reported that $\mu(n)$ is nearly the same for electrons and holes \cite{Nimtz1983}. At low temperatures, the long-range Coulomb potential scattering has been suggested as the dominant limiting factor, which leads to an enhancement in $\mu$ with increasing $n$ due to enhanced screening \cite{Ravich1971a}. In low-dimensional heterostructures, however, additional scattering mechanisms become relevant, including background and remote impurity scattering, interface roughness, and alloy disorder \cite{Prinz1999,Huang2022a,Chung2022}. 

The evolution of $1/eR_{\mathrm{H}}$ extracted from Hall measurements is consistent with the changes observed in the SdH oscillation patterns (Fig.~\ref{fig:RvsB}c), confirming controllable shifts of $E_{\mathrm{F}}$ under PPC gating. The quantized Hall plateaus observed in the highest fields correspond to filling factors $\nu = 3$ for holes and $\nu = 1$ for electrons \cite{Krizman2024,Kazakov2025}. These values coincide with the degeneracies of the $\bar{M}$ and $\bar{\Gamma}$ valleys, respectively (Fig.~\ref{fig:bands}a). The results, therefore, provide direct evidence of a photoinduced transition from a threefold-degenerate $\bar{M}$-valley two-dimensional hole gas (2DHG) to a single $\bar{\Gamma}$-valley polarized two-dimensional electron gas (2DEG).

\begin{figure}
    \centering
    \includegraphics[width=1.00\columnwidth]{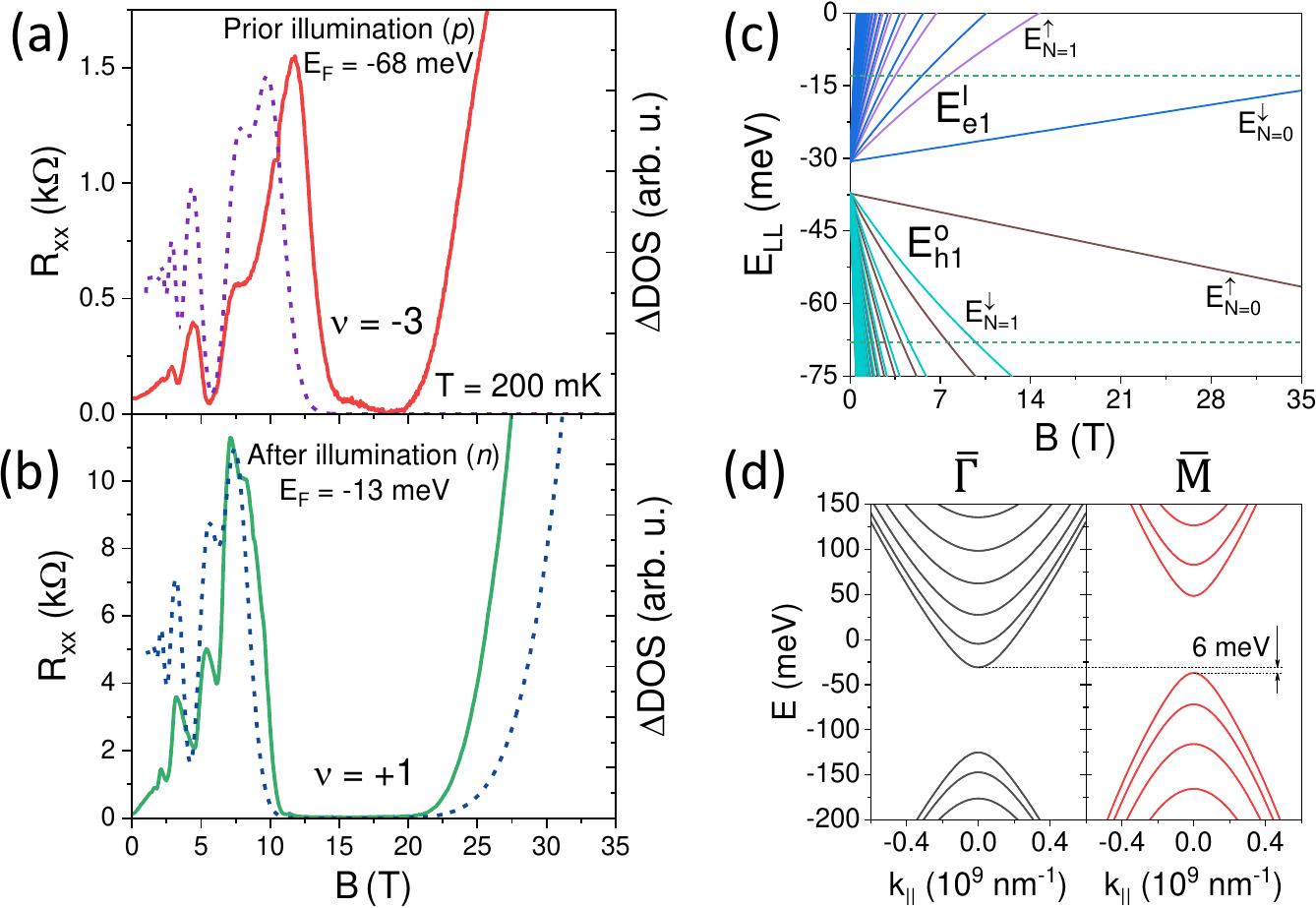}
    \caption{
    (a,b) Longitudinal resistance $R_{\mathrm{xx}}$ measured in magnetic fields up to 35~T (solid lines) demonstrates robust quantum Hall states in both the {\it p}-type (a) and {\it n}-type (b) regimes, characterized by nearly vanishing $R_{\mathrm{xx}}$ on the corresponding plateaus.
    These states are well reproduced by the \kp model calculations (dashed lines) of the density-of-states (DOS) oscillations and corresponding LL fan diagram (c). Green dashed lines indicate the Fermi levels corresponding to the {\it p}- and {\it n}-type quantum Hall states shown in panels (a) and (b).
    (d) Calculated subband dispersions for the longitudinal $\bar{\Gamma}$ and oblique $\bar{M}$ valleys.
    }
    \label{fig:kp}
\end{figure}

An abrupt increase of the resistance on approaching the CNP from both sides signalizes a transition to the strongly localized regime at the CNP vicinity, which sets-in at carrier densities of about $2\times10^{11}$\,cm$^{-2}$ per valley and is accompanied by an increase of resistivity by a factor of 1.7 between 20 and 2 K at CNP \cite{sup}. Quantum localization by short-range scattering, driven by the Anderson and Altshuler-Aronov mechanisms, has been invoked to describe localization in semiconductors with a large magnitude of the static dielectric constant and absence of highly-localized deep carrier traps \cite{Prinz1999}. According to another scenario, transitions from the strongly to weakly localized regime occur at the onset of the percolation transport over charge puddles created by potential fluctuations in compensated narrow-gap semiconductors \cite{Huang2022b}. The calculated dispersion of the longitudinal and oblique subbands indicates a small indirect band gap of approximately $\sim$6~meV (Fig.~\ref{fig:kp}d), much smaller than the Fermi energy at $2\times10^{11}$~cm$^{-2}$.

A closer inspection of the low-field magnetoresistance reveals the significance of quantum localization phenomena, as negative magnetoresistance attains 20\% (Fig.~\ref{fig:RvsB}d). Previous magnetoresistance studies on thin films of Pb$_{1-x}$Sn$_x$Se grown on SrTiO$_{3}$ (111) substrate showed a complex competition between weak localization (WL) and weak antilocalization (WAL) depending on $x_{\mathrm{Sn}}$, film thickness, and temperature \cite{Zhang2015a}. In all samples studied here, we consistently observe a robust positive magnetoresistance at very small fields, characteristic of WAL \cite{sup} and indicative of significant spin-orbit coupling in the system \cite{Turowski2023}. In MR data, this feature is most pronounced near the CNP, where a sharp WAL cusp appears around magnetic fields on the order of 10~mT. As the field increases above 100~mT, the positive MR is gradually replaced by a negative contribution, indicative of WL. Away from the CNP, both WAL and WL signatures become dominated by temperature-independent positive parabolic magnetoresistance, specific to multivalley systems.

When expressed in terms of the magnetoconductivity, $\Delta\sigma_{\mathrm{xx}}$ (Fig.~\ref{fig:RvsB}e,f), the WAL feature is found to be practically insensitive to the position of the Fermi level. In multivalley systems such as (Pb,Sn)Se QWs, intervalley scattering is expected to influence the amplitude of quantum interference effects \cite{Nestoklon2014}. In the regime where the intervalley scattering time is shorter than the phase-coherence time, contributions from different valleys are effectively mixed, and the system behaves as a single coherent channel, consistent with our observations. Notably, even in epilayers containing a larger number of interface defects \cite{Kazakov2021}, the WAL amplitude is well described by the Hikami–Larkin–Nagaoka (HLN) model with a prefactor $\alpha=0.5$ \cite{Hikami1980}, supporting the dominance of strong intervalley scattering. Moreover, considering possible effects of Eu spins in the barriers, we note that according to experimental and theoretical studies of the exchange integrals \cite{Dietl1994}, we expect stronger spin-disorder scattering by Eu spins for holes than for electrons in the non-topological range of Sn contents. Within such a scenario, spin-disorder scattering reduces the magnitude of WAL and diminishes the rise of WL magnetoresistance in the case of holes.


High magnetic field measurements at $T=200$~mK reveal well-developed quantum Hall states in both the {\it n}- and {\it p}-type regimes at temperatures down to $\sim$200~mK. Figure~\ref{fig:kp} presents magnetotransport data for sample B in fields up to 35~T. In each regime, illumination-induced tuning of the $E_{\mathrm{F}}$ enables access to the lowest Landau level (LL), producing pronounced Hall plateaus accompanied by vanishing longitudinal resistance. Upon further increasing the magnetic field, QHE states give way to a high non-saturating MR. These observations demonstrate that the quantum Hall effect persists after photodoping and can be effectively controlled via the PPC state, preserving the key features previously reported in strain-engineered (Pb,Sn)Se QWs \cite{Krizman2024}. The high quality of the QHE results reconfirms the absence of parallel conductance across the barriers at low temperatures and under PPC conditions. We also note that the persistence of the PPC effect in high magnetic fields indicates that exchange coupling of carriers with paramagnetic Eu spins, and resulting phenomena such as bound magnetic polarons, are not essential for the existence of the PPC.

To interpret these observations, we performed \kp-model calculations using the framework established in the previous work \cite{Kazakov2025}. In the present calculations, we adjusted the remote band contributions by changing their signs in the $g$-factors to reduce their absolute values, following \cite{Pascher1988}, thereby enabling higher accuracy. The resulting model accurately reproduces both the low-field SdH oscillations and the observed quantized Hall plateaus. The dashed curves in Figs.~\ref{fig:kp}(a,b) represent the calculated density-of-states modulation, which aligns closely with the resistance extrema in $R_{\mathrm{xx}}$. The width of the lowest LL plateau in the {\it n}-type regime likewise matches the theoretical results. 

\subsection{Comparison between electrostatic and optical gating}

Finally, we compare optical and electrostatic gating as complementary approaches to tuning $E_{\mathrm{F}}$, using sample D as a representative case. Previous studies on HgTe QWs \cite{Shuvaev2022} and (Bi,Sb)$_{2}$Te$_{3}$ thin films \cite{Yeats2015} have reported no qualitative distinction between optical and electrostatic gating, whereas in InAs/GaSb bilayers these two methods were shown to induce distinctly different electronic configurations near the CNP \cite{Meyer2024}.

\begin{figure}
    \centering
    \includegraphics[width=1.00\columnwidth]{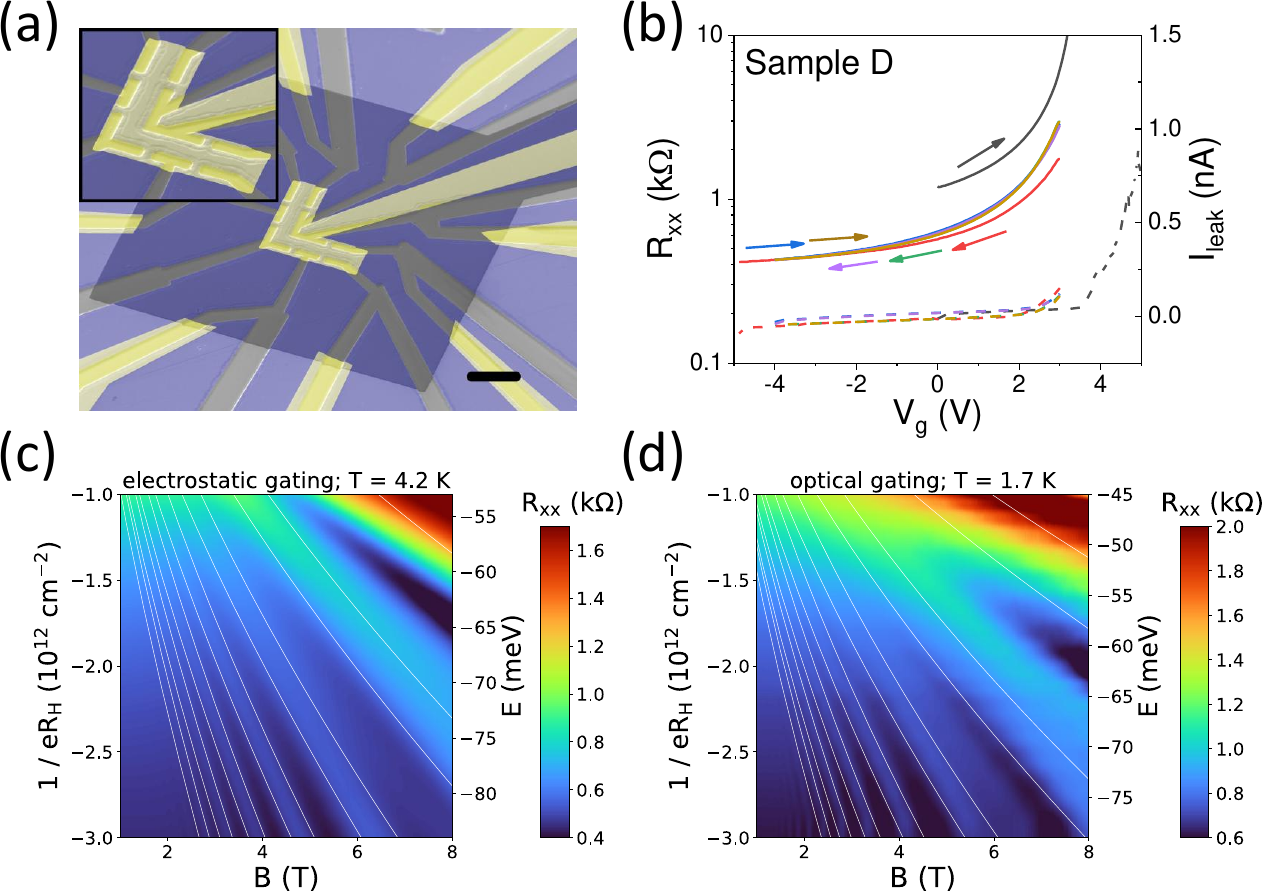}
    \caption{
    (a) False-color SEM image of the FET device fabricated from sample D, viewed at a tilt angle of 44$^{\circ}$. Gold (yellow) denotes the gate electrode and contacts, the insulating HfO$_{2}$ layer is shown in shaded grey, the 2D hole gas is highlighted in light violet, and the etched region appears in grey. The inset shows an L-shaped Hallbar. The black scale bar corresponds to 20~µm.
    (b) Gate-voltage dependence of the longitudinal resistance $R_{\mathrm{xx}}$. An initial sweep up to +5~V induces gate leakage and leads to pronounced hysteresis. Beyond this initial cycle, subsequent sweeps within the range –5~V to +3~V exhibit reversible characteristics.
    (c,d) Evolution of resistance SdH oscillations as the Fermi level is tuned via electrostatic gating (c) and optical gating (d). In both cases, the SdH oscillations align with the calculated Landau level spectrum, as highlighted by the overlaid white lines. The right axes indicate the corresponding Landau level energies.}
    \label{fig:fet}
\end{figure}

To explore this question, we fabricated a field-effect transistor (FET) device in an L-shaped Hall-bar geometry (Fig.~\ref{fig:fet}(a)), incorporating a 100~nm HfO$_{2}$ dielectric grown by atomic layer deposition and capped with a 100~nm Au gate. Upon initial sweeping of the gate voltage $V_{\mathrm{g}}$ to +5 V, the longitudinal resistance $R_{\mathrm{xx}}$ increased beyond 10~k$\Omega$, eventually entering a non-Ohmic regime. When subsequently reduced to +3 V, the system did not return to its initial state, indicating strong hysteresis. Following this initial sweep, subsequent gating cycles within the voltage range of –5~V to +3~V displayed negligible hysteresis, although within this window the hole density could not be reduced much below $\sim-10^{12}$~cm$^{-2}$. That gives a limitation, similar to noted above, of $\sim-3\cdot10^{11}$~cm$^{-2}$ holes per valley. The FET hysteresis effect is usually associated with charge trapping/detrapping processes at insulator-material interfaces \cite{Hinz2006,Lunczer2019}. Though the source of hysteresis in the device under study is not known precisely, its presence is also consistent with the observation of the PPC effect as discussed below.

Remarkably, a similar limitation was encountered in the ungated Hall-bar device: the optical gating could only tune the system down to a comparable hole density of $\sim 10^{12}$ cm$^{-2}$. Near this density threshold, the recombination dynamics became comparable to the magnetotransport measurement timescale, preventing further approach to the CNP. In a separate spectral response study, while illuminating with photon energy that provides more efficient electron pumping, the sample eventually also entered a non-ohmic regime \cite{sup}. A comparison of carrier-density-dependent SdH oscillations in electrostatically and optically gated configurations [Fig.~\ref{fig:fet}(c,d)] reveals an identical oscillation sequence in both cases, each in excellent agreement with the calculated LL spectrum. These findings indicate that, within the accessible doping range, optical gating produces an equivalent electronic configuration to electrostatic gating, while offering a simpler and low-cost alternative for initial assessment of QW properties.

\subsection{Spectral response of the PPC}

\begin{figure*}
    \centering
    \includegraphics[width=1.8\columnwidth]{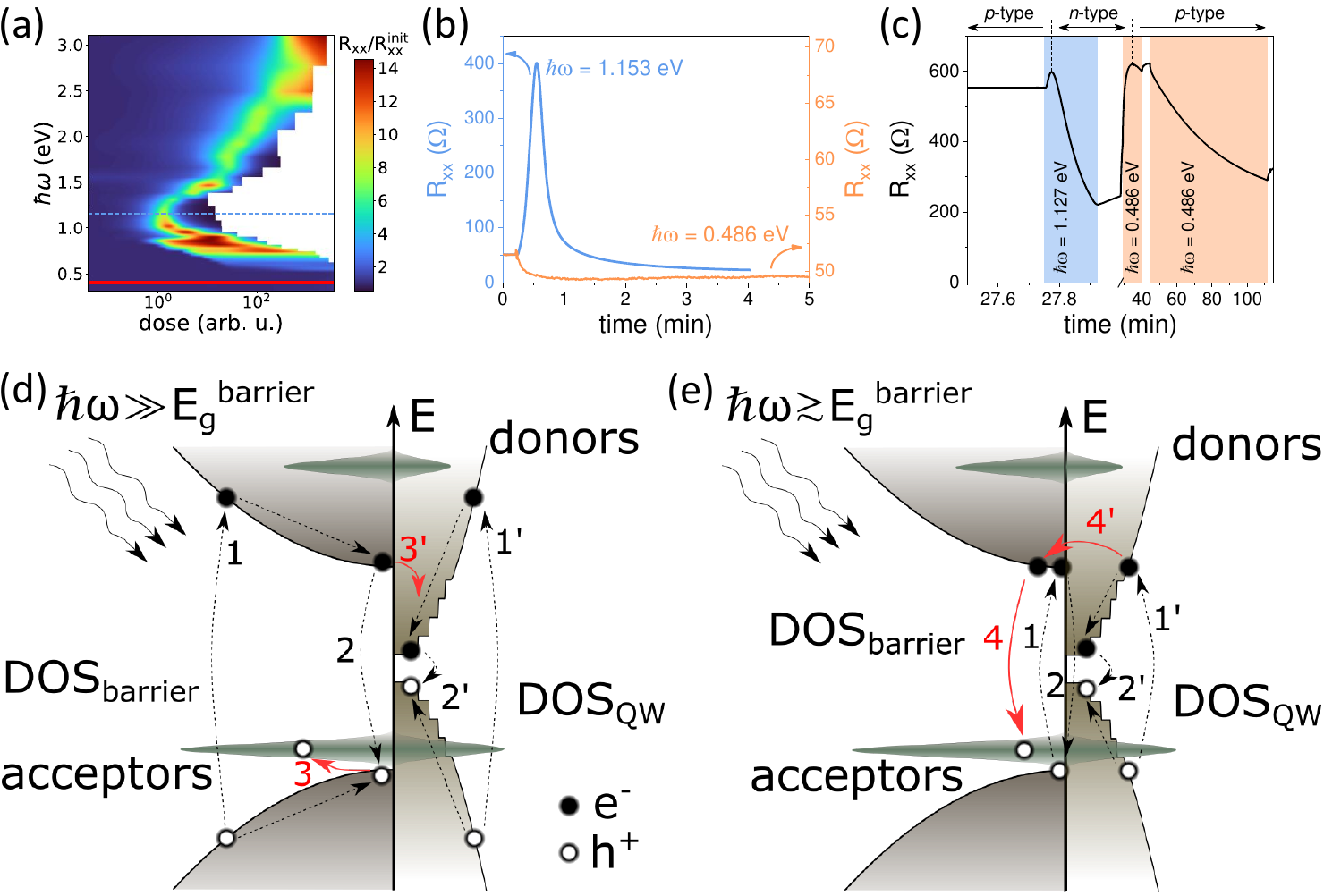}
    \caption{
    (a) Spectral response of the photoconductivity, plotted as a function of the energy of the incident photon and dose (measured in arb.~u. because of an unknown geometrical factor). The data is normalized to the initial value of the QW resistance. Dashed lines indicate wavelengths for temporal dependencies in (b). A thick red line indicates the bandgap of the (Pb,Eu)Se barriers. 
    (b) Two examples of the temporal dependence of the photoconductivity, showing both positive and negative effects. The dark drift in the resistivity value due to slow thermalisation was subtracted for clarity. 
    (c) Part of the temporal dependence, which shows the effect of the subsequent illumination of the QW with two wavelengths corresponding to positive and negative photoconductivities.
    (d, e) Schematic representation of the proposed PPC mechanism. Photons excite electrons and holes in the barrier and QW regions (processes 1 and 1' respectively), and part of those pairs quickly recombine (processes 2 and 2'). For the photon energies greater than the barrier bandgap, $\hbar\omega>E_{\mathrm{g}}^{\mathrm{barrier}}$ (d), process 1 generates more photoelectrons than process 1' due to the larger joint DOS. Part of those photoelectrons escape the barrier to the QW (process 3'), while photoholes get trapped in the acceptor band (process 3). For the $\hbar\omega\gtrsim E_{\mathrm{g}}^{\mathrm{barrier}}$ (e), process 1' takes over and generates more photoelectrons in the QW than in the barrier. Those photoelectrons partially penetrate the barrier (process 4'), where they recombine with trapped holes (process 4). 
    }
    \label{fig:spectra}
\end{figure*}

To clarify the nature of the PPC effect in the studied QWs, we examined the spectral photoresponse of the $R_{\mathrm{xx}}$ over a broad wavelength range of $\lambda=400-5000$~nm, which corresponds to photon energies $\hbar\omega=0.25-3.1$~eV. For particular $\hbar\omega$ values, the temporal evolution $R_{\mathrm{xx}}(t)$ under illumination was recorded. To allow a direct comparison between different excitation energies, the illumination time $t$ was converted into a photon dose $d$, defined as $d\propto{Pt}/{\hbar\omega}$, where $P$ is a measured power at a given $\hbar\omega$. Thus, $d$ is proportional to the number of incident photons, up to a geometrical factor arising from the different positioning and sizes of the sample and power meter. The results obtained for sample B are presented in Fig.~\ref{fig:spectra}a; qualitatively similar behavior was observed for all investigated samples \cite{sup}.

For all studied QWs, the strongest photoresponse occurs for $\hbar\omega\approx$1.2~eV. At both higher and lower excitation energies, a larger photon dose is required to induce an equivalent change in resistance, although the qualitative evolution of $R_{\mathrm{xx}}(t)$ remains similar. When the photon energy decreases below $\approx$~0.3--0.4~eV, no measurable photoresponse is observed. This energy range approximately coincides with the energy gap of the QW barriers (indicated by the thick red line in Fig.~\ref{fig:spectra}a). As the excitation energy approaches this threshold from above, a weak positive photoconductivity becomes visible, i.e., resistance {\it diminishes} under illumination (Fig.~\ref{fig:spectra}b).

Such behavior indicates that long-wavelength IR illumination reverses carrier density changes generated by high-energy photons. To test this hypothesis, we alternated the illumination between $\hbar\omega$=1.127~eV and 0.486~eV, the resistance exhibits alternating peaks corresponding to repeated passages through the CNP, as illustrated in Fig.~\ref{fig:spectra}c and \cite{sup}. Note that when light was switched off, the relaxation process has opposite trend, indicating that light heating was not responsible for PPC reversal.

\subsection{Origin of the PPC in (Pb,Sn)Se/(Pb,Eu)Se quantum wells}
\label{sec:mechanism}


The PPC effect in bulk semiconductors has usually been assigned to a negative correlation energy $U$ of carriers bound to relevant impurities, associated with either lattice distortion (e.g., DX centers in III–V and II–VI semiconductors \cite{Mooney1990,Wojtowicz1993,Semaltianos1993}) or a specific electronic configuration (e.g., PbTe:Ga \cite{Belogorokhov2000,Volkov2002}). Typically, PPC increases the electron and hole concentration in $n$-type and $p$-type materials, respectively \cite{Lany2005}. In our $p$-type QWs, the illumination {\it reduces} the hole concentration and can persistently turn the QW into the $n$-type at low temperatures. 

The absence of any measurable photoresponse for $\hbar\omega<E^{\mathrm{barrier}}_{\mathrm{g}}$ indicates that the barriers play a crucial role in the PPC mechanism in our QWs. The observed efficiency of the PPC effect is centred at $\hbar\omega\approx1.1–1.2$~eV in all samples. Considering joint density of states for the barrier valence and conduction bands, and the photon penetration depth, this value corresponds to the highest carrier generation rate in the barrier region near the interface (transitions 1 in Fig.~\ref{fig:spectra}d). Furthermore, this energy corresponds also to the expected position of the Eu $4f^{6/7}$ state with respect to the barrier conduction band \cite{Kowalski1999}, which contains Eu $5d$ orbitals, allowing for dipole optical transitions. Meanwhile, photons also generate photocarriers in the QW (transitions 1' in Fig.~\ref{fig:spectra}d). Considering joint density of states, it is clear that at $\hbar\omega$ approaching $E^{\mathrm{barrier}}_{\mathrm{g}}$, the process 1' generates more photocarriers at energies corresponding to barrier band edges than the process 1 (Fig.~\ref{fig:spectra}e).

Altogether, the experimental results indicate that photoelectrons generated in the barrier conduction band are transferred to the QW (process 3', Fig.~\ref{fig:spectra}d). This accumulation of photoelectrons in the QW is persistent, which means that their recombination with photoholes is blocked at low temperatures. At $\hbar\omega$ approaching $E^{\mathrm{barrier}}_{\mathrm{g}}$ transfer of photoelectrons from the barrier region to the QW (process 3', Fig.~\ref{fig:spectra}d) is overruled by the reverse process, i.e. by the transfer of photoelectrons from the QW to the barrier (process 4', Fig.~\ref{fig:spectra}e). Such a transfer ceases to be possible for $\hbar\omega<E^{\mathrm{barrier}}_{\mathrm{g}}$ as photoelectrons do not have high enough energy to penetrate the barriers, and fast recombination between all photoelectrons and photoholes occurs within the QW (process 2', Fig.~\ref{fig:spectra}e). 

We assign the asymmetry between behaviors of photoelectrons and photoholes to a different energy position of acceptors (hole traps) and compensating donors (electron traps) with respect to the band edges in PbSe. As shown in Fig.~\ref{fig:spectra}d,e, the acceptor band, presumably originating from Pb vacancies, resides close to the top of the valence band, whereas the donor band, presumably brought about by anion vacancies, is high in the conduction band. Such positions are consistent with the higher achievable electron concentrations compared to the hole concentrations in undoped PbSe, and explain also a possibility of obtaining insulating PbSe (our sample A at low temperatures) and the activation energy of conductance, $E_d = 88$\,meV, in Pb$_{0.89}$Eu$_{0.11}$Se (see Supplemental Material (SM) \cite{sup}). This model also indicates that the acceptors in barriers act as remount dopants, so that the 3D hole concentration is larger in narrower (Pb,Sn)Se QW, as observed (see Table I and Ref.\,\cite{Kazakov2025}). We recall at this point that carriers bound to the defect states we discuss here are highly localized, whereas textbook shallow levels associated with charged defects (such as an empty donor state resonant with the conduction band), due to large dielectric constant and low carrier masses in IV-VI compounds, are mixed up with the band states and, thus, irrelevant here. 

Now, we see that photoholes in the barriers are quickly trapped by unoccupied acceptor states that exist due to compensation (process 3, Fig.~\ref{fig:spectra}d). Their low-temperature hopping toward the interface is hampered by high spatial localization, dispersion in the acceptor energies, and formation of lattice polarons or AX$^{+}$ centres. In contrast, the lifetime of photoelectrons in the conduction band is longer, so that some of them reach the interface and are captured by the QW. Accordingly, after turning the light off, we have a surplus of both electrons in the QW and immobile holes in the semi-insulating barriers. A temperature increase activates hopping of holes toward the interface, where they recombine with electrons whose wave functions penetrate the barriers, the process leading to equalization of the Fermi levels in the QW and barriers. If $\hbar\omega$ approaches $E^{\mathrm{barrier}}_{\mathrm{g}}$, the generation of photocarriers in the QW takes over. Again, photoholes are captured by acceptor traps in the QW, whereas a part of the photoelectrons diffuses to the barriers and recombine with resident holes (processes 4 and 4', Fig.~\ref{fig:spectra}e). 

\section{Summary and Outlook}

In this work, we have demonstrated that the PPC effect in (Pb,Sn)Se/(Pb,Eu)Se QWs can be employed as an efficient optical gate, capable of continuously tuning the $E_{\mathrm{F}}$ and even switching the conductivity type from {\it p}-type to {\it n}-type. This transition is confirmed by the sign inversion of the low-field Hall response and a change in quantum Hall plateau degeneracy, reflecting a photoinduced crossover from a threefold-degenerate $\bar{M}$-valley two-dimensional hole gas to a $\bar{\Gamma}$-valley polarized two-dimensional electron gas. Since Bi act as an efficient donor in the narrow-gap (Pb,Sn)Se QWs, it can set the initial position of the $E_{\mathrm{F}}$ \cite{Volobuev2017}. That opens a way to engineer the photoresponse of the QWs, which can range from positive (samples A, B, C) to negative photoconductivity (samples B, C, D). Also, a photoinduced transition from an insulating to a conducting regime can be realized (sample A). Furthermore, in samples B and C, the strain-induced energy offset between longitudinal and oblique valleys enables optical control over the valley occupancy, effectively realizing valley polarization tuning through PPC.

Analysis of the spectral dependence of the $R_{\mathrm{xx}}(t)$ response reveals not only the optimal photon energies that induce the PPC state, but also those that can reverse it, demonstrating the reversibility of the effect. Additionally, the spectral response and properties of the PPC effect lead us to conclude that the asymmetry in the locations of the acceptor and donor defect states accounts for photoinduced electron doping in the (Pb,Sn)Se QWs. As a result, the PPC carrier-density modulation closely mimics conventional electrostatic gating. Overall, our results highlight the broad versatility of optical gating as an efficient and low-cost technique for controlling conductivity type and valley population in (Pb,Sn)Se-based QWs.

While electrostatic gating remains the conventional method for $E_{\mathrm{F}}$ tuning, PPC-induced modulation of conductivity offers unique advantages for programmable and spatially reconfigurable patterning of {\it p-n} junctions, as demonstrated in other systems \cite{Yeats2015}. These optically written junctions can be erased by local heating, high currents, or IR radiation, enabling reversible device architectures. Moreover, extending this concept to the quantum Hall regime opens new possibilities for QHE-based {\it p-n} junctions \cite{Williams2007,Abanin2007}, enriched by the additional control over the valley degree of freedom. These findings position IV–VI PPC-engineered heterostructures as a promising platform for reconfigurable quantum Hall devices and optically programmable valleytronics, as well as a promising platform for the higher Chern number QAHE system \cite{Fang2014,Majewski2026}. We expect that the thermal stability of the QAHE can be improved by the location in the exchange gap of a level with a short carrier localization length, for instance, the defect state observed here in (Pb,Eu)Se or the Cr impurity level investigated recently in (Pb,Sn)Te \cite{Krolicka2025}.

\section*{Acknowledgments}

This research was partially supported by the Foundation for Polish Science project ”MagTop” no. FENG.02.01-IP.05-0028/23 co-financed by the European Union from the funds of Priority 2 of the European Funds for a Smart Economy Program 2021-2027 (FENG) 
and by Narodowe Centrum Nauki (NCN, National Science Centre, Poland) IMPRESS-U Project No. 2023/05/Y/ST3/00191. 
V.V.V. also acknowledges long-term program of support of the Ukrainian research teams at the Polish Academy of Sciences carried out in collaboration with the U.S. National Academy of Sciences with the financial support of external partners. 
Measurements at high magnetic fields were supported by LNCMI-CNRS, members of the European Magnetic Field Laboratory (EMFL) and by the Ministry of Education and Science, Poland (grant no. DIR/WK/2018/07) via its membership to the EMFL. 
This research was supported by funds from the state budget allocated by the Minister of Science (Polska) as part of the Polish Metrology II programme project no. PM-II/SP/0012/2024/02 amount of funding 944,900.00 PLN, total project value PLN 944,900.00. 

Studied heterostructures were grown by G.K., V.V.V. and G.S.; A.K. carried out microstructure processing and magnetotransport measurements; M.S., W.W. and A.K. measured spectral dependencies; C.-W.C. and B.A.P. assisted in high-field magnetotransport measurements. T.~Wojciechowski assisted in microstructure processing. A.K. and T.D. performed \kp calculations. T.~Wojtowicz and T.D. were responsible for funding acquisition and general management. The manuscript was written by A.K. and T.D. with input from all coauthors. All authors discussed the results and commented on the manuscript.

\section*{Data availability}

The data that support the findings of this article are openly available \cite{zenodo}.

\bibliography{bibliography}

\end{document}


\title{\thetitle}

\author{Alexander~Kazakov\orcidA}
\email{kazakov@MagTop.ifpan.edu.pl}
\affiliation{International Research Centre MagTop, Institute of Physics, Polish Academy of Sciences, Aleja Lotnikow 32/46, PL-02668 Warsaw, Poland}
\author{Gauthier~Krizman\orcidB}
\affiliation{Laboratoire de Physique de l’Ecole normale sup{\'e}rieure, ENS, Universit{\'e} PSL, CNRS, Sorbonne Universit{\'e}, 24 rue Lhomond 75005 Paris, France}
\author{Valentine~V.~Volobuev\orcidH}
\affiliation{International Research Centre MagTop, Institute of Physics, Polish Academy of Sciences, Aleja Lotnikow 32/46, PL-02668 Warsaw, Poland}
\affiliation{National Technical University "KhPI", Kyrpychova Str. 2, 61002 Kharkiv, Ukraine}
\author{Micha\l~Szot\orcidJ}
\affiliation{International Research Centre MagTop, Institute of Physics, Polish Academy of Sciences, Aleja Lotnikow 32/46, PL-02668 Warsaw, Poland}
\author{Wojciech~Wo{\l}kanowicz\orcidK}
\affiliation{International Research Centre MagTop, Institute of Physics, Polish Academy of Sciences, Aleja Lotnikow 32/46, PL-02668 Warsaw, Poland}
\author{Chang-Woo~Cho\orcidC}
\affiliation{Laboratoire National des Champs Magn{\'e}tiques Intenses, CNRS, LNCMI, Universit{\'e} Grenoble Alpes, Univ. Toulouse, INSA Toulouse, EMFL, F-38042 Grenoble, France}
\affiliation{Department of Physics, Chungnam National University, Daejeon, 34134, Republic of Korea}
\affiliation{Institute for Sciences of the Universe, Chungnam National University, Daejeon 34134, South Korea}
\author{Benjamin~A.~Piot\orcidD}
\affiliation{Laboratoire National des Champs Magn{\'e}tiques Intenses, CNRS, LNCMI, Universit{\'e} Grenoble Alpes, Univ. Toulouse, INSA Toulouse, EMFL, F-38042 Grenoble, France}
\author{Tomasz~Wojciechowski\orcidE}
\affiliation{International Research Centre MagTop, Institute of Physics, Polish Academy of Sciences, Aleja Lotnikow 32/46, PL-02668 Warsaw, Poland}
\author{Gunther~Springholz\orcidF}
\affiliation{Institut f{\"u}r Halbleiter- und Festk{\"o}rperphysik, Johannes Kepler University, Altenbergerstrasse 69, A-4040 Linz, Austria}
\author{Tomasz~Wojtowicz\orcidG}
\affiliation{International Research Centre MagTop, Institute of Physics, Polish Academy of Sciences, Aleja Lotnikow 32/46, PL-02668 Warsaw, Poland}
\author{Tomasz~Dietl\orcidI}
\email{dietl@MagTop.ifpan.edu.pl}
\affiliation{International Research Centre MagTop, Institute of Physics, Polish Academy of Sciences, Aleja Lotnikow 32/46, PL-02668 Warsaw, Poland}

\onecolumngrid
\renewcommand{\thefigure}{S\arabic{figure}}
\renewcommand{\thetable}{S\arabic{table}}
\renewcommand{\theequation}{S\arabic{equation}}
\renewcommand{\thepage}{S\arabic{page}}
\renewcommand{\thesection}{S\arabic{section}}

\setcounter{page}{1}
\setcounter{section}{0}
\setcounter{equation}{0}
\setcounter{figure}{0}
\setcounter{table}{0}

\begin{center}
\textbf{\Large Supplemental Material} \\
\vspace{0.2in} \textmd{\Large \thetitle}\\
\end{center}

\section{Energy profile of the studied samples}

\begin{figure}[h]
    \centering
    \includegraphics[width=0.6\columnwidth]{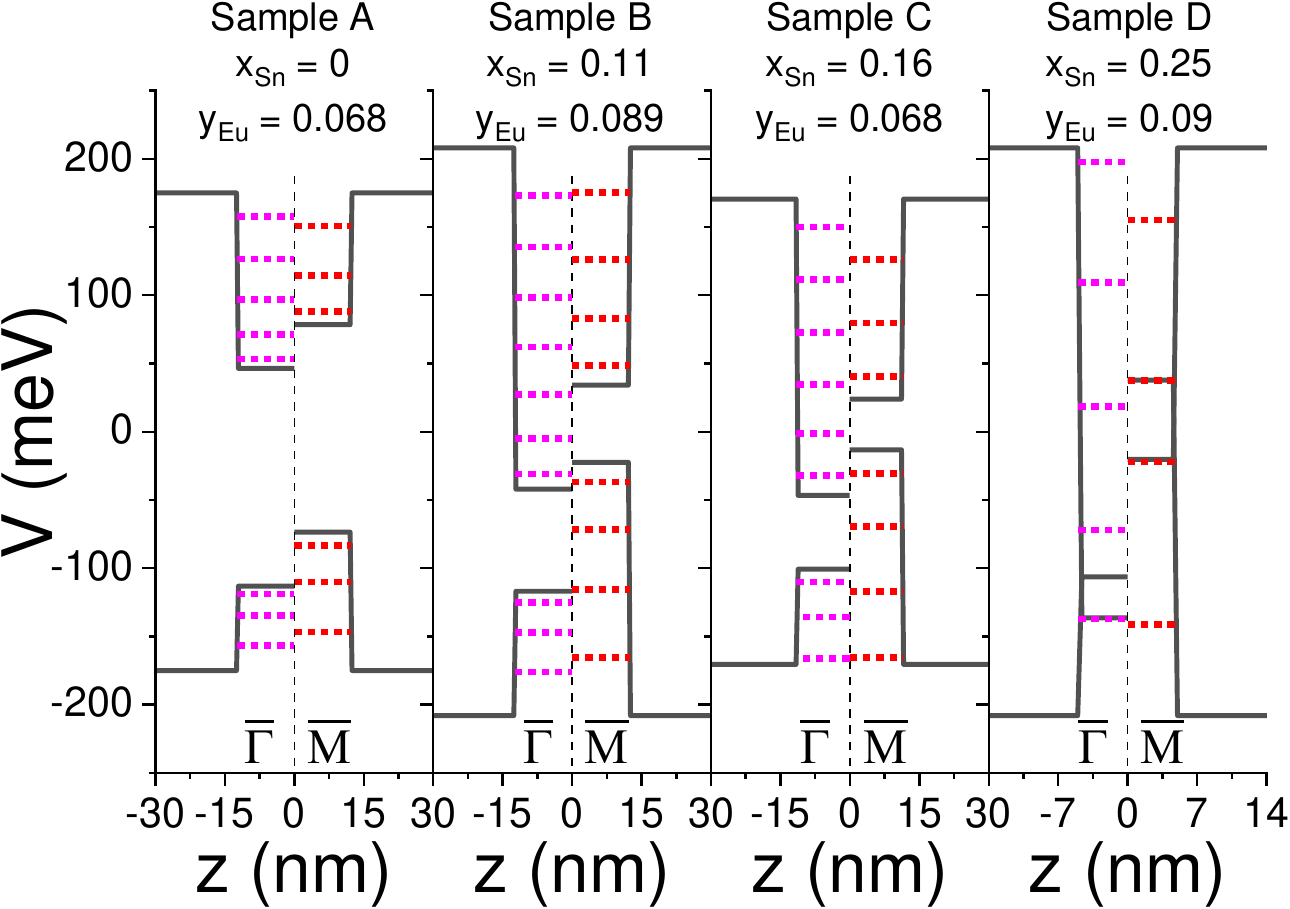}
    \caption{
    The gaps and subbands in the studied samples for the longitudinal and oblique valleys. QW levels were calculated using the \kp model. Only sample D has a bulk inverted gap, though the QW is still in the trivial regime. For each QW, the left side depicts the position of the subbands for the single longitudinal $\bar{\Gamma}$-valley, while the left -- for the triple-degenerated oblique $\bar{M}$-valleys.
    }
    \label{fig:QW_subbands}
\end{figure}

\section{Conductivity of barrier material as a function of temperature}

\begin{figure}[h]
    \centering
    \includegraphics[width=0.4\columnwidth]{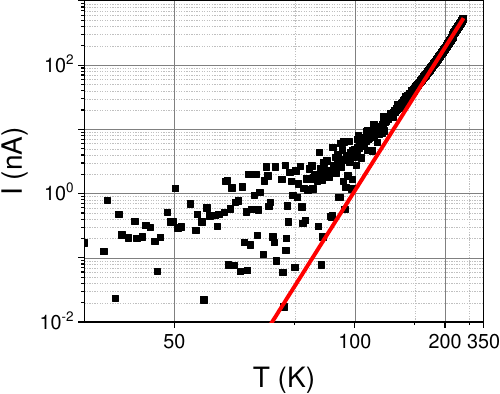}
    \caption{
    Conductivity of the barrier material. For measurements, a 1~$\mu$m film of Pb$_{0.89}$Eu$_{0.11}$Se was grown on (111) BaF$_{2}$ substrate. Conductivity was measured using soldered In contacts in a 2-terminal mode, under dc 100~mV bias. Red line is the activation function $I\propto\exp(-\frac{E_d}{k_BT})$ fitting, which gives $E_d=88$~meV.
    }
    \label{fig:barrier_conduct}
\end{figure}

\section{Temperature dependence as a function of the $E_{\mathrm{F}}$}

\begin{figure}[h]
    \centering
    \includegraphics[width=0.8\columnwidth]{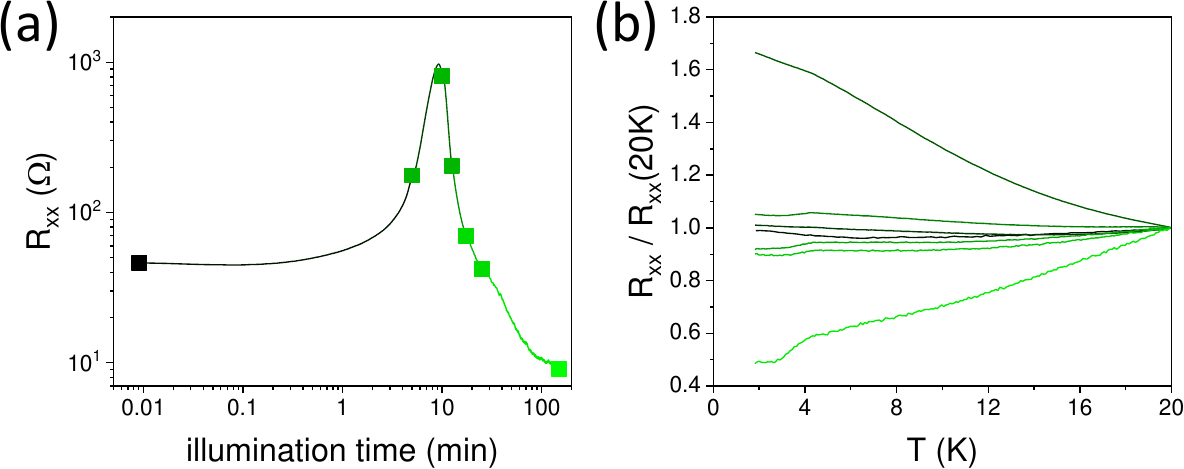}
    \caption{
    (a) Variation of the sample B longitudinal resistance $R_{\mathrm{xx}}$ as a function of time under the action of LED illumination. At several points (marked), the temperature dependence of resistance was measured, which is depicted in (b). At the CNP, we observe a semiconducting type of the $R_{\mathrm{xx}}(T)$ dependence, while far away from the CNP, it is metallic.
    }
    \label{fig:sup_3645_RvsT}
\end{figure}

\section{Optical pumping pace as a function of the $I_{\mathrm{LED}}$}

\begin{figure}[h]
    \centering
    \includegraphics[width=0.8\columnwidth]{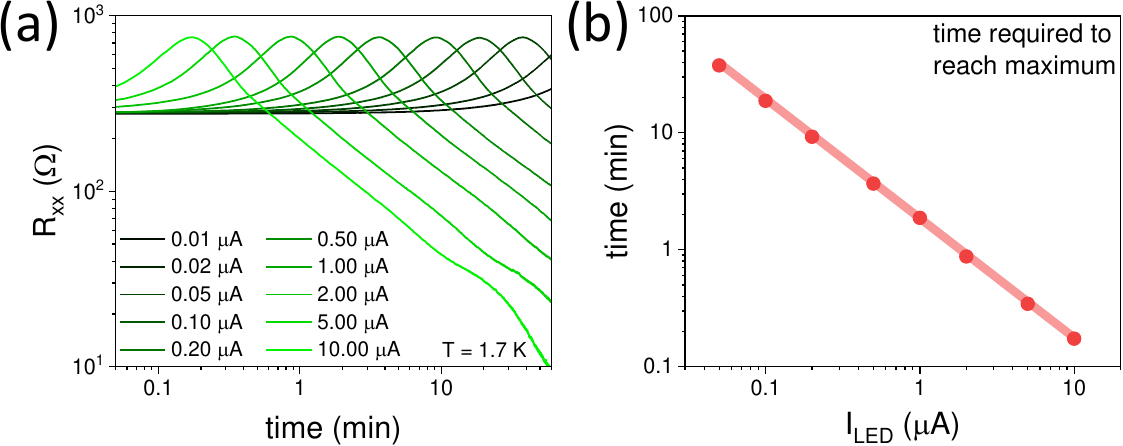}
    \caption{
    (a) Variation of the sample B longitudinal resistance $R_{\mathrm{xx}}$ with time under LED illumination of various intensities. LED intensity was varied by passing different values of the current $I_{\mathrm{LED}}$.
    (b) Characteristic time $t$ of the optical gating was chosen as the time required to reach the maximum resistivity. Fitting $t(I_{\mathrm{LED}})$ (line) yields $t\propto\frac{1}{I_{\mathrm{LED}}}$, indicating that electron pumping speed is proportional to the light intensity.
    }
    \label{fig:sup_3695_ILED}
\end{figure}

\section{Full time dependence}

\begin{figure}[h]
    \centering
    \includegraphics[width=\columnwidth]{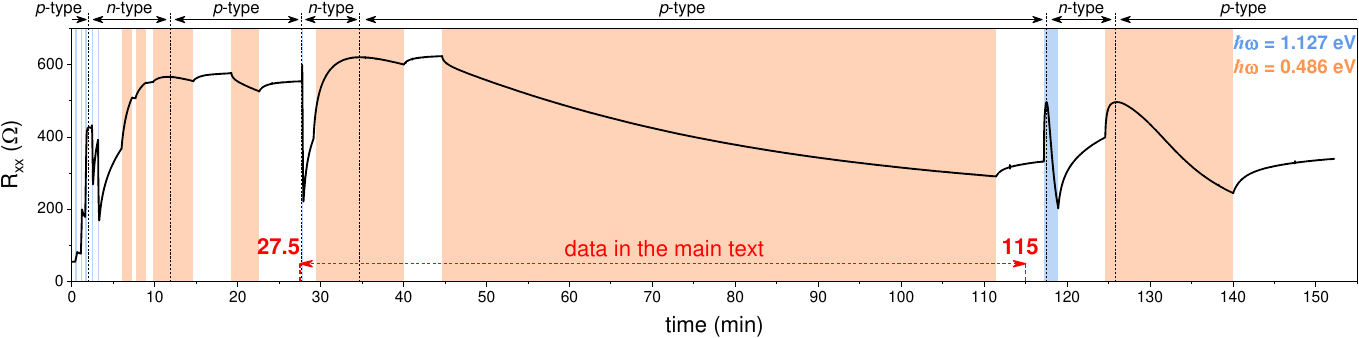}
    \caption{
    Full-time dependence, displayed in Fig.~\ref{fig:spectra}c. The blue and orange shaded areas indicate illumination periods. 
    }
    \label{fig:sup_3645_fulltime_dep}
\end{figure}

\clearpage
\newpage

\section{Additional data on other studied samples}

\subsection{Sample A}

\begin{figure}[h]
    \centering
    \includegraphics[width=\columnwidth]{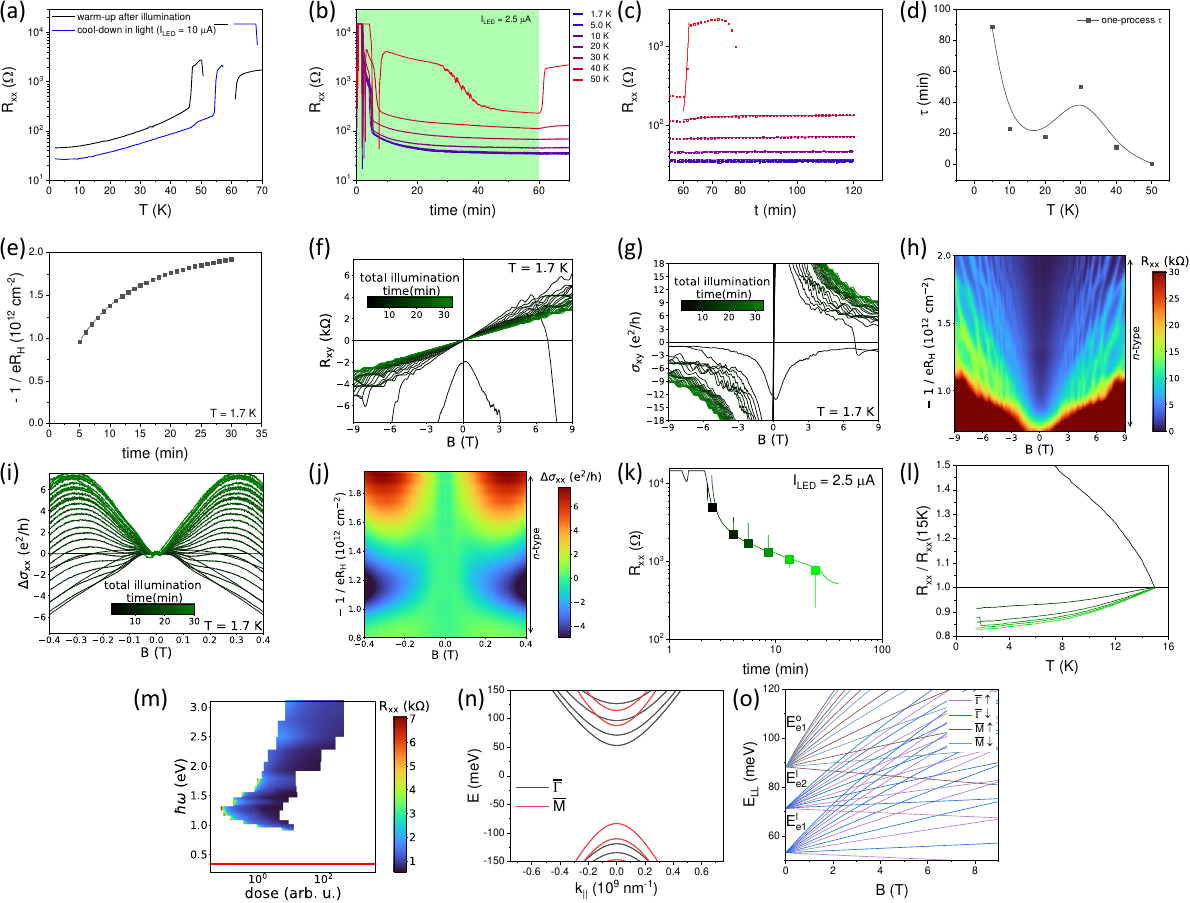}
    \caption{{\bf Extended set of data on sample A.}
    (a) Comparison of $R_{\mathrm{xx}}(T)$ curves between two measurement protocols: warming up after illuminating at low temperatures (black), and cooling down while illuminating (blue). There is no cool-down in the dark, since sample A was not conducting before illumination.
    (b) Dynamics of longitudinal resistance under illumination at different temperatures.
    (c) $R_{\mathrm{xx}}(t)$ after shutting down illumination, fitting with an exponential relaxation function (solid lines).
    (d) Dependence of the relaxation time versus temperature, obtained from the fitting results.
    (e) Dependence of the low-field inverse Hall coefficient on the total illumination time.
    (f, g) Evolution of the $R_{\mathrm{xy}}$ and $\sigma_{\mathrm{xy}}$ with the total illumination time.
    (h) Evolution of the $R_{\mathrm{xx}}$ and SdH oscillations with the low-field Hall coefficient.
    (i, j) Low-field part of the $\sigma_{\mathrm{xx}}$ as a function of the total illumination time and inverse Hall coefficient, showing that WAL feature is practically independent of the position of the $E_{\mathrm{F}}$.
    (k) $R_{\mathrm{xx}}(t)$, measured at low temperatures with points at which $R_{\mathrm{xx}}(T)$ in (l) were taked.
    (m) Photon energy dependence of the PPC effect with red line indicating $E_{\mathrm{g}}^{\mathrm{barrier}}$ value.
    (n, o) Results of the \kp calculations, showing subband dispersion and LLs close to the conduction band edge for both longitudinal $\bar{\Gamma}$ and oblique $\bar{M}$ valleys. 
    }
    \label{fig:sup_3773_relax}
\end{figure}

\clearpage
\newpage
\subsection{Sample C}

\begin{figure}[h]
    \centering
    \includegraphics[width=\columnwidth]{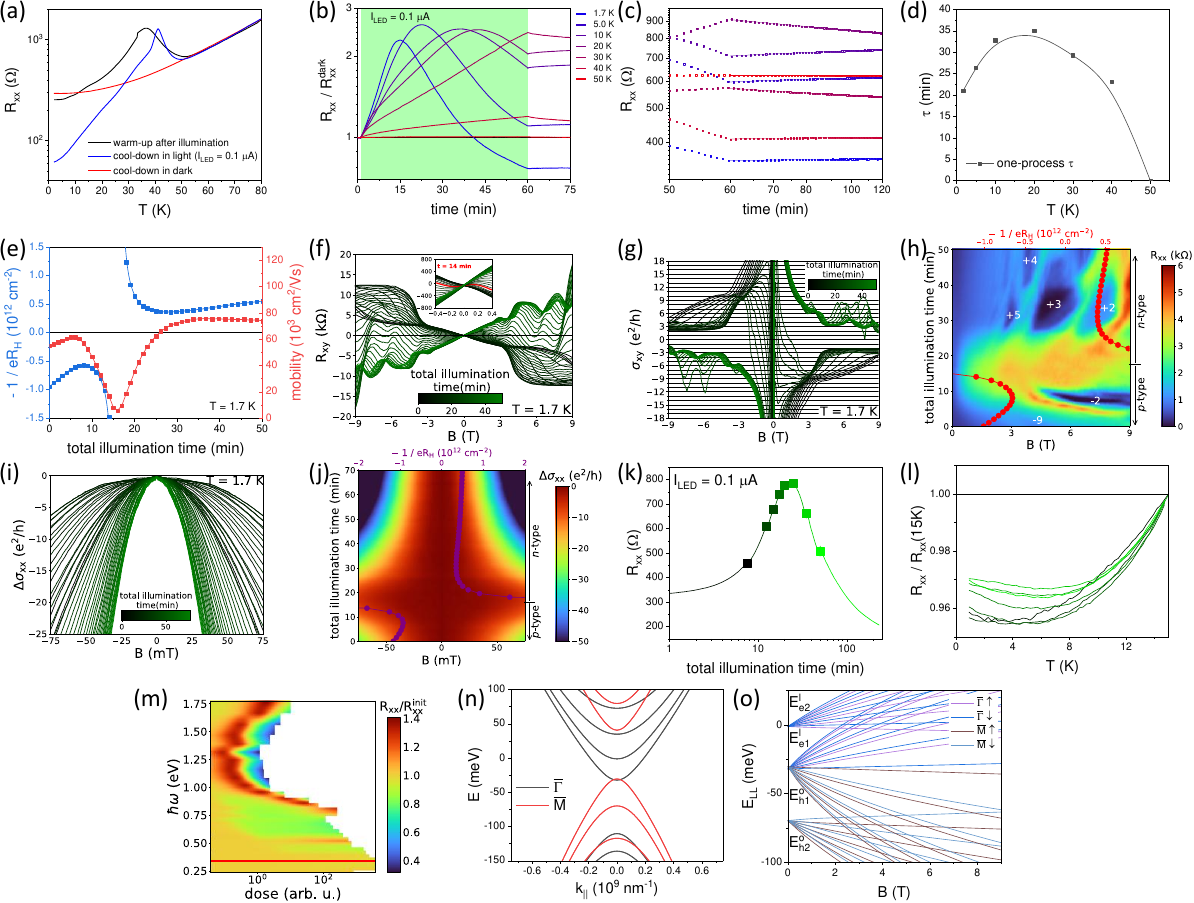}
    \caption{{\bf Extended set of data on sample C.}
    (a) Comparison of $R_{\mathrm{xx}}(T)$ curves between three measurement protocols: warming up after illuminating at low temperatures (black); cooling down in dark (blue); and cooling down while illuminating (red).
    (b) Dynamics of longitudinal resistance under illumination at different temperatures.
    (c) $R_{\mathrm{xx}}(t)$ after shutting down illumination, fitting with an exponential relaxation function (solid lines).
    (d) Dependence of the relaxation time versus temperature, obtained from the fitting results.
    (e) Dependence of the low-field inverse Hall coefficient and mobility on the total illumination time.
    (f, g) Evolution of the $R_{\mathrm{xy}}$ and $\sigma_{\mathrm{xy}}$ with the total illumination time. The red curve in the inset shows non-linear behavior, indicating mixed-type conductivity in the vicinity of the CNP.
    (h) Evolution of the $R_{\mathrm{xx}}$ and SdH oscillations with the total illumination time. Red dots indicate the corresponding low-field Hall coefficient.
    (i, j) Low-field part of the $\sigma_{\mathrm{xx}}$ as a function of the total illumination time and inverse Hall coefficient, showing that WAL feature is practically independent of the position of the $E_{\mathrm{F}}$.
    (k) $R_{\mathrm{xx}}(t)$, measured at low temperatures with points at which $R_{\mathrm{xx}}(T)$ in (l) were taked.
    (m) Photon energy dependence of the PPC effect with red line indicating $E_{\mathrm{g}}^{\mathrm{barrier}}$ value.
    (n, o) Results of the \kp calculations, showing subband dispersion and LLs close to the conduction band edge for both longitudinal $\bar{\Gamma}$ and oblique $\bar{M}$ valleys.
    }
    \label{fig:sup_3695_relax}
\end{figure}

\clearpage
\newpage
\subsection{Sample D}

\begin{figure}[h]
    \centering
    \includegraphics[width=\columnwidth]{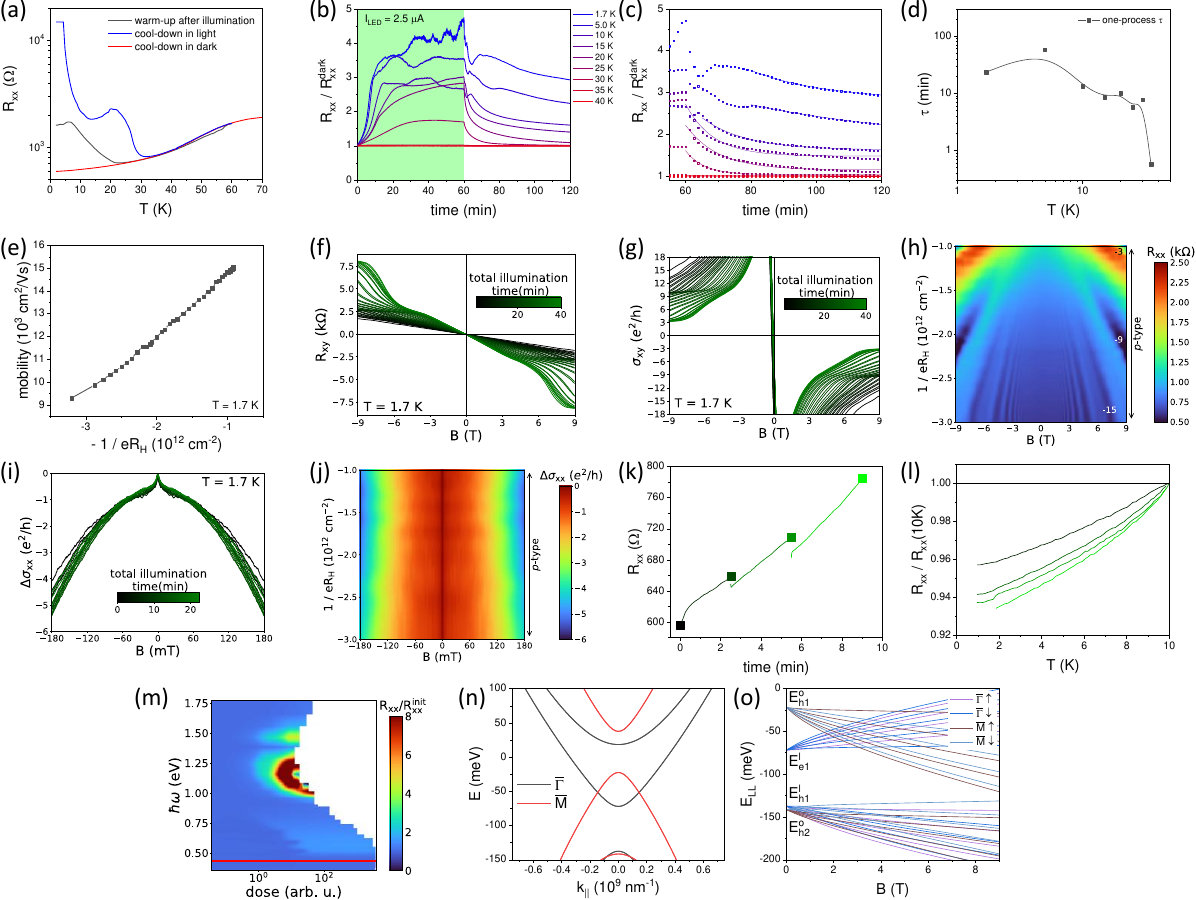}
    \caption{{\bf Extended set of data on sample D.}
    (a) Comparison of $R_{\mathrm{xx}}(T)$ curves between three measurement protocols: warming up after illuminating at low temperatures (black); cooling down while illuminating (blue); and cooling down in dark (red).
    (b) Dynamics of longitudinal resistance under illumination at different temperatures.
    (c) $R_{\mathrm{xx}}(t)$ after shutting down illumination, fitting with an exponential relaxation function (solid lines).
    (d) Dependence of the relaxation time versus temperature, obtained from the fitting results.
    (e) Dependence of the mobility on the low-field inverse Hall coefficient. Points correspond to different total illumination times.
    (f, g) Evolution of the $R_{\mathrm{xy}}$ and $\sigma_{\mathrm{xy}}$ with the total illumination time.
    (h) Evolution of the $R_{\mathrm{xx}}$ and SdH oscillations with the low-field Hall coefficient.
    (i, j) Low-field part of the $\sigma_{\mathrm{xx}}$ as a function of the total illumination time and inverse Hall coefficient, showing that WAL feature is practically independent of the position of the $E_{\mathrm{F}}$.
    (k) $R_{\mathrm{xx}}(t)$, measured at low temperatures with points at which $R_{\mathrm{xx}}(T)$ in (l) were taked.
    (m) Photon energy dependence of the PPC effect with red line indicating $E_{\mathrm{g}}^{\mathrm{barrier}}$ value. At $\hbar\omega\approx1.2$~eV, the sample eventually ended in a very high resistance state, where contacts lost ohmicity.
    (n, o) Results of the \kp calculations, showing subband dispersion and LLs close to the conduction band edge for both longitudinal $\bar{\Gamma}$ and oblique $\bar{M}$ valleys.
    }
    \label{fig:sup_3411_relax}
\end{figure}